\documentclass[
  aps,
  prb,
  reprint,
  amsmath,
  amssymb,
  longbibliography
]{revtex4-2}

\usepackage{graphicx}
\usepackage{dcolumn}
\usepackage{bm}

\usepackage[T1]{fontenc}

\usepackage{orcidlink}

\hypersetup{
  colorlinks=true,
  citecolor=blue,
  linkcolor=black,
  urlcolor=black
}

\newcommand*{\CV}{C_{V}}
\newcommand{\Debye}{{D_3}}

\newcommand*{\EDebye}{E_{\scriptscriptstyle \mathrm{D}}}
\newcommand*{\Estatic}{E_{\mathrm{s}}}
\newcommand*{\Ezero}{E_{0}}

\newcommand*{\eostag}{\text{\tiny EOS}}
\newcommand*{\gammaeos}{\gamma_{\eostag}}
\newcommand*{\gammaeosinf}{\gamma_{\infty}^{\eostag}}

\newcommand*{\gammainf}{\gamma_{\infty}}
\newcommand*{\gammazero}{\gamma_{0}}
\newcommand*{\gtwo}{g_{2}}

\newcommand*{\kB}{k_{\scriptscriptstyle \mathrm{B}}}

\newcommand*{\Kprime}{K^{\prime}}
\newcommand*{\Kprimezero}{K^{\prime}_{0}}
\newcommand*{\Kstatic}{K_{\mathrm{s}}}
\newcommand*{\Kstaticprime}{K_{\mathrm{s}}^{\prime}}

\newcommand*{\Kadiabatic}{K_{S}}
\newcommand*{\Kisothermal}{K_{T}}
\newcommand*{\Kzero}{K_{0}}
\newcommand*{\Kzerocalc}{K_{0}^{\mathrm{calc}}}
\newcommand*{\Kzeroexpt}{K_{0}^{\mathrm{expt}}}

\newcommand*{\PFGzero}{P_{\mathrm{FG},0}}
\newcommand*{\Pstatic}{P_{\mathrm{s}}}

\newcommand{\RKRT}{\mathcal{R}_K^{\mathrm{RT}}}

\newcommand*{\Tzero}{T_{0}}

\newcommand*{\thetaD}{\Theta_{\scriptscriptstyle \mathrm{D}}}
\newcommand*{\thetaDzero}{\Theta_{0}}

\newcommand*{\teffzero}{t_{\mathrm{eff},0}}
\newcommand*{\tone}{t_{1}}
\newcommand*{\tzero}{t_{0}}

\newcommand*{\Ve}{V_{e}}
\newcommand*{\Vzero}{V_{0}}
\newcommand*{\Vzerocalc}{V_{0}^{\mathrm{calc}}}
\newcommand*{\Vzeroexpt}{V_{0}^{\mathrm{expt}}}

\begin{document}

\title{Finite-temperature bulk moduli from an EOS-based Gr\"uneisen function}
\author{ \c{C}etin K{\i}l{\i}\c{c} \orcidlink{0000-0003-2690-4940}}
\affiliation{
Department of Physics, Gebze Technical University, Gebze, Kocaeli 41400, T\"{u}rkiye
}

\date{\today}

\begin{abstract}
An equation-of-state (EOS)-based construction of the Gr\"uneisen function is developed and assessed through predictions of finite-temperature bulk moduli. 
In this approach, the volume dependence of the Gr\"uneisen function is expressed analytically in terms of static EOS information, 
anchored by Debye temperatures obtained from elastic data sampled near equilibrium, and constrained by the infinite-compression limit. 
The resulting form requires as material-specific input only static energy-volume data and near-equilibrium elastic properties, with no parameters adjusted to thermal data. 
Within a Mie--Gr\"uneisen--Debye framework, the approach is examined for
diamond, magnesium oxide, silicon, and sodium chloride, chosen to span a broad
range of stiffness, using machine-learning interatomic potentials from the
Universal Models for Atoms (UMA) and Universal Point Edge Transformer (UPET)
families, together with a dispersion-corrected variant of UMA.
The calculated bulk moduli reproduce the expected experimental softening trends and capture the overall scale of the bulk-modulus temperature derivatives, 
although the level of quantitative agreement depends on the material, the underlying interatomic potential, and the selected analytic EOS form.
These results show that an EOS-based construction of the Gr\"uneisen function can capture the leading thermal-softening behavior of bulk moduli without fitting to thermal data. 
The observed sensitivity to the underlying static description further suggests that the framework may help identify deficiencies relevant to thermoelastic transferability and thereby inform future training and validation strategies for universal machine-learning interatomic potentials.
\end{abstract}

\maketitle

\section{\label{sec:intro}INTRODUCTION}

The Gr\"uneisen parameter $\gamma$ links microscopic lattice excitations to the macroscopic thermoelastic response of a solid and remains a key quantity in finite-temperature condensed-matter physics.
Its volume dependence, $\gamma(V)$, strongly affects thermal pressure, thermal expansion, and the temperature dependence of elastic moduli. 
Accordingly, many thermodynamic models depend sensitively on the assumed form of $\gamma(V)$ \cite{ZharkovKalinin1971,anderson1995equations}.

The Gr\"uneisen function $\gamma(V)$ can be obtained from phonon spectra or elastic data computed for compressed and expanded volumes, 
or from analytical expressions involving equation-of-state (EOS) quantities. 
Phonon- and elasticity-based approaches are closely connected to microscopic physics, 
but in practice they may become numerically fragile because of imaginary phonon frequencies, proximity to mechanical instability, 
and noise in higher-order derivatives of the total energy. 
By contrast, EOS-based approaches yield smooth trends at modest computational cost, 
but they are often only weakly constrained and may therefore provide an inadequate description of $\gamma(V)$ 
\cite{ZharkovKalinin1971,StaceyHodgkinson2019}.

A generalized EOS-based expression for the Gr\"uneisen function was formulated by Burakovsky and Preston \cite{burakovsky2004analytic}:
\begin{equation}
\gammaeos(V)
=\frac{\frac{1}{2}\Kstaticprime-\frac{1}{6}-\frac{1}{3}t\!\left(1-\frac{\Pstatic}{3\Kstatic}\right)+\frac{\Pstatic}{3\Kstatic}\,V\frac{dt}{dV}}
{1-2t\,\frac{\Pstatic}{3\Kstatic}},
\label{eq:gamma_eos}
\end{equation}
where $t=t(V)$ is an analytic function of volume,
$\Pstatic$ denotes the static pressure,
$\Kstatic$ is the corresponding bulk modulus,
and $\Kstaticprime$ is its first pressure derivative, all evaluated along the static isotherm.
In the constant-$t$ limit, Eq.~\eqref{eq:gamma_eos} reduces to the
Slater \cite{Slater1939},
Dugdale--MacDonald \cite{dugdale1953thermal},
Vashchenko--Zubarev \cite{VashchenkoZubarev1963},
and Barton--Stacey \cite{barton1985gruneisen, stacey1995theory}
relations corresponding to $t=0,1,2$, and 2.35, respectively.
Equation~\eqref{eq:gamma_eos} also recovers the Irvine--Stacey form \cite{irvine1975pressure}
under the reparameterization introduced in Ref.~\onlinecite{burakovsky2004analytic}.

\begin{figure}
  \includegraphics[width=0.47\textwidth]{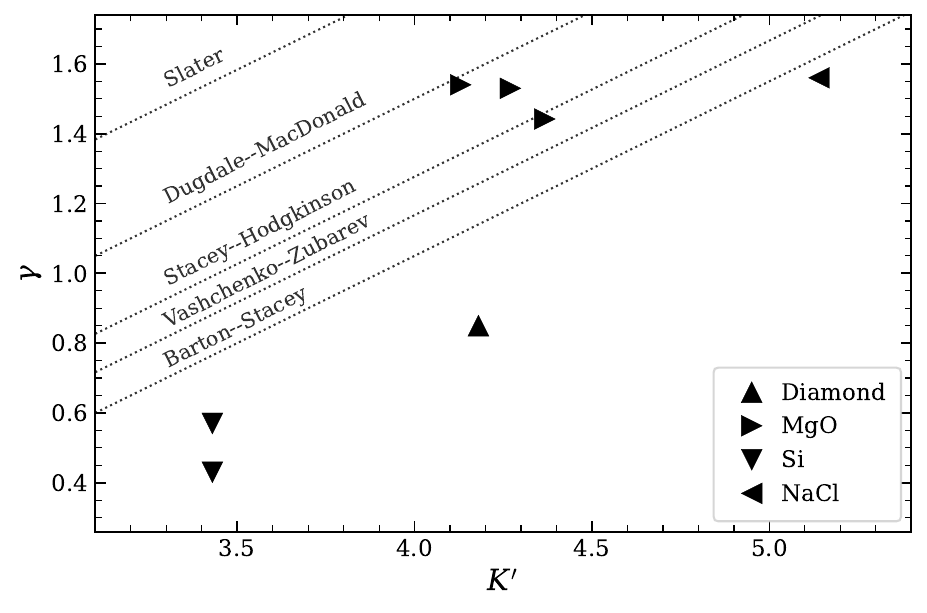}
\caption{
Experimental values of the Gr\"uneisen parameter $\gamma$ plotted against the pressure derivative of the bulk modulus $\Kprime$ for
MgO \cite{Kono2010PETI, Tange2009, isaak1989measured},
NaCl \cite{matsui2012simultaneous},
diamond \cite{Dewaele2008High},
and Si \cite{gschneidner1964physical, Gauster1971, anzellini2019quasi}.
The dotted reference lines are drawn using the zero-pressure limit of Eq.~\eqref{eq:gamma_eos},
$\gamma = \Kprime/2 - 1/6 - t/3$,
with $t=0,1,1.67,2$, and 2.35 corresponding to the Slater \cite{Slater1939}, Dugdale--MacDonald \cite{dugdale1953thermal}, 
Stacey--Hodgkinson \cite{StaceyHodgkinson2019}, Vashchenko--Zubarev \cite{VashchenkoZubarev1963}, and Barton--Stacey \cite{barton1985gruneisen, stacey1995theory} relations, respectively.
}
  \label{fig:exptgammavsKprime}
\end{figure}

The constant-$t$ approach requires selecting a reasonable value of $t$,
also denoted by $m$ or $f$ in the literature \cite{ZharkovKalinin1971,anderson1995equations},
but that choice is not straightforward and can lead to substantial deviations from experiment.
For example, Stacey and Hodgkinson adopted the empirical value $f=1.67$ in their analysis of lower-mantle data for the Earth \cite{StaceyHodgkinson2019},
rather than one of the aforementioned theoretical values.
More generally, adopting any single constant-$t$ prescription can be problematic,
as illustrated in Fig.~\ref{fig:exptgammavsKprime}, where representative ambient values of $\gamma$ and $\Kprime$ are compared with the corresponding reference lines.
The spread of the data shows that a universal constant-$t$ choice does not provide a uniformly satisfactory description across the materials examined here.

These considerations motivate a generalized framework for constructing $\gamma(V)$ within Eq.~\eqref{eq:gamma_eos}, 
in which $t$ varies with volume, or equivalently with pressure along the EOS path. 
Variable-$t$ formulations have previously been developed primarily 
for all-density analytic representations of the cold Gr\"uneisen parameter and related quantities, 
with experimental compression or melting information used as input \cite{burakovsky2004analytic,burakovsky2004cold}. 
The present approach instead introduces a system-specific construction of $\gamma(V)$ 
in which the volume dependence is expressed analytically from static-lattice EOS information, 
anchored by the elasticity-derived equilibrium Gr\"uneisen parameter, 
and constrained by the infinite-compression limit.
The objective is to obtain a physically motivated and minimally parameterized form of $\gamma(V)$ 
requiring only static energy--volume data and near-equilibrium elastic properties, 
both derived from atomistic calculations, without fitting any parameters to thermal data. 
In the present work, the atomistic calculations are carried out using machine-learning interatomic potentials (MLIPs) from the Universal Models for Atoms \cite{wood2025uma} and Universal Point Edge Transformer \cite{pushing-unconstrained-2026} families; however, the same framework is directly applicable to density-functional theory (DFT) calculations and empirical interatomic potentials.

The practical workflow is organized accordingly.
Static energy--volume $E_{\mathrm{s}}(V)$ data are generated from constrained relaxations at fixed volume and fitted to either the Vinet \cite{vinet1987universal} or AP2 \cite{Holzapfel1998} EOS.
Elastic stiffness tensors are then evaluated at several volumes in the near-equilibrium region, 
from which Debye temperatures are obtained and used to anchor the equilibrium value of the Gr\"uneisen function.
An infinite-compression constraint, required only for the Vinet EOS, is subsequently imposed to complete the construction.
Together, the static EOS information, near-equilibrium elastic data, and, where required, the asymptotic constraint define the methodological basis of the present approach.

The approach is assessed through predictions of the finite-temperature bulk modulus $K(T)$
within the Mie--Gr\"uneisen--Debye framework.
The bulk modulus is the natural immediate target because it governs the volumetric response and enters directly into the thermal-pressure formulation; extension to other elastic moduli, such as the shear modulus and Young's modulus, is reserved for a future tensorial thermoelastic treatment.
The analysis is carried out for diamond, MgO, Si, and NaCl,
chosen to span a broad range of stiffness from very hard to comparatively soft solids.
The computed bulk-modulus curves are compared with available experimental trends to determine 
whether the proposed EOS-based construction of $\gamma(V)$ reproduces the principal thermal softening trends 
and the overall scale of the bulk-modulus temperature derivative, $dK/dT$, without fitting to thermal data.
Although experimental determinations of $K(T)$ are available for selected materials, systematic datasets spanning broad temperature ranges and well-defined thermodynamic conditions remain limited and scattered across the literature.
For this reason, developing complementary and computationally efficient workflows for predicting $K(T)$ in a thermodynamically consistent manner, without fitting to experimental thermal data, is of clear practical importance, especially for newly synthesized or hypothetical materials for which such data are sparse or unavailable.

It should be noted that an effective Gr\"uneisen function can, in principle, be extracted from \emph{ab initio} molecular-dynamics (AIMD) simulations without fitting to experimental thermal data, 
as demonstrated, for example, in Ref.~\onlinecite{Oganov2003}. 
In practice, however, AIMD-based thermal-EOS treatments commonly introduce an assumed functional form for $\gamma(V)$ \cite{Oganov2003} to parametrize its volume dependence and 
thereby avoid performing an independent AIMD simulation at every $(V,T)$ or $(P,T)$ state point. 
Such direct sampling is computationally demanding because AIMD calculations require sufficiently large supercells, long equilibration and sampling times, and substantial computational resources \citep{Zhang2023}.

The present approach is not intended to replace more comprehensive finite-temperature methods, such as AIMD, quasiharmonic lattice dynamics, and thermodynamic integration, 
when their additional physical detail is required and sufficient computational resources are available \citep{Karki2001,Gillan2006,Wentzcovitch2010Thermoelasticity}. 
Rather, the aim is to develop a computationally inexpensive route in which $\gamma(V)$ is inferred from static EOS data and near-equilibrium elastic information, 
without direct sampling of finite-temperature configurations. 
The resulting workflow can thus serve as a rapid screening and diagnostic tool by testing whether a given static description produces physically plausible thermal trends 
before more expensive finite-temperature simulations are undertaken. 
This role remains relevant when MLIPs are used in place of direct first-principles calculations, 
because the same workflow can reveal deficiencies in the static EOS and near-equilibrium elastic response, 
as illustrated below for the MLIPs considered in this work. 
This sensitivity of the predicted thermal behavior to these static ingredients provides a physically motivated diagnostic of thermoelastic transferability 
and may thereby inform future training and validation strategies for universal MLIPs.

The remainder of this paper is organized as follows.
Section~\ref{s:met} describes the present approach and summarizes the computational details.
Section~\ref{s:res} presents and discusses the results.
Finally, concluding remarks are given in Sec.~\ref{s:con}.

\section{\label{s:met}METHODS}

In this section, the present approach and the workflow for implementing it are described.
First, the forms adopted for the volume dependence of $\gamma(V)$ and $t(V)$ are specified, together with the static EOS and near-equilibrium elastic input used to anchor them, as well as the means used to impose the infinite-compression limit.
Next, the resulting $\gamma(V)$ is employed within the Mie--Gr\"uneisen--Debye framework to compute the temperature dependence of the bulk modulus.
Finally, the MLIP-based atomistic workflow for generating the static energy--volume data and near-equilibrium elastic properties is summarized.

\subsection{EOS-based construction of the Gr\"uneisen function with elastic anchoring}

Although a unique functional form for $t(V)$ in Eq.~\eqref{eq:gamma_eos} is not known \emph{a priori},
Ref.~\onlinecite{burakovsky2004analytic} suggests an analytic structure in which $t$ is expanded in powers of $V^{1/3}$.
In this work, a lowest-order truncation of that expansion is adopted, expressed in dimensionless form as
\begin{equation}
t(V) = \tzero - \tone \left( \frac{V}{\Vzero} \right)^{1/3},
\label{eq:tV}
\end{equation}
where $\Vzero$ denotes the static equilibrium volume, and $\tzero$ and $\tone$ are constants.
Higher-order terms would provide additional flexibility, but their coefficients would not be uniquely determined by the asymptotic and equilibrium constraints employed here and would require additional independent constraints, microscopic input, or fitting to thermal data.
Analysis in Ref.~\onlinecite{burakovsky2004analytic}
indicates that choosing $\tzero = 5/2$ yields the infinite-compression limit $\gammainf = 1/2$,
as discussed there for a lattice of positive ions embedded in a degenerate electron gas.
Accordingly, the present approach fixes $\tzero = 5/2$, and hence $\gammainf = 1/2$.
In constant-$t$ constructions, by contrast, it is more common to adopt the degenerate-electron-gas value $\gammainf = 2/3$, e.g., Ref.~\onlinecite{holzapfel2001equations}.
With $\tzero$ fixed by the asymptotic constraint, the remaining parameter $\tone$
is determined by the equilibrium anchoring condition, as described below.

Whereas the present construction fixes $\gammainf$ at a theoretically motivated value,
$\gammainf$ is often treated as an unconstrained parameter in fits to experimental thermodynamic data,
e.g., in Refs.~\cite{Dorogokupets2007,Tange2009},
yielding substantially larger values that are attributable to the fact that the fitted data remain far from the infinite-compression regime and therefore do not directly constrain the asymptotic limit.
For the present construction, the more consequential issue is that
evaluating Eq.~\eqref{eq:gamma_eos} for a given EOS in the infinite-compression limit yields
$\gammaeosinf$, which may differ substantially from $\gammainf$.
For the Vinet EOS, $\gammaeosinf=1/6$, whereas the AP2 EOS gives
$\gammaeosinf=1/2=\gammainf$.
When required, a constant shift $s=\gammainf-\gammaeosinf$ is therefore applied:
\begin{equation}
\gamma(V)=\gammaeos(V)+s.
\label{eq:gamma}
\end{equation}
Accordingly, $s=1/3$ for Vinet but $s=0$ for AP2.
The volume-independent shift does not alter the underlying EOS and therefore
does not address potential limitations associated with extrapolating analytic EOS forms
to extreme compressions beyond their typical ranges of validity
\cite{holzapfel1996physics}.
It does, however, enforce the imposed asymptotic value as $V \to 0$.
EOS forms for which $\gammaeosinf$ substantially exceeds $\gammainf$ require particular scrutiny,
since the resulting negative shift may drive $\gamma(V)$ to unphysical values even at moderate compressions.

A further difficulty arises when an EOS reduces to a lower-order form at a particular parameter value.
For example, the third-order Birch--Murnaghan (BM3) EOS \cite{birch1947finite} yields $\gammaeosinf=4/3$ and hence $s=-5/6$ for $\Kprimezero\neq4$.
At $\Kprimezero=4$, however, the third-order term vanishes and BM3 reduces to BM2, for which $\gammaeosinf=1$ and $s=-1/2$.
Thus, for materials with $\Kprimezero$ close to 4, the choice of asymptotic shift becomes ambiguous within the Birch--Murnaghan framework.

The parameter $\tone$ is determined by imposing the equilibrium anchoring condition
$\gamma(\Vzero)=\gammazero$, where $\gammazero$ denotes the Gr\"uneisen parameter at the static equilibrium volume.
Using $\Pstatic(\Vzero)=0$ in Eq.~\eqref{eq:gamma_eos} yields
$\gammaeos(\Vzero)=\Kprimezero/2-1/6-(\tzero-\tone)/3$, and therefore
\begin{equation}
\tone = 3\left(\gammazero + \frac{1}{6} + \frac{\tzero}{3} - \frac{\Kprimezero}{2} - s\right),
\label{eq:tone}
\end{equation}
where $\Kprimezero$ is shorthand for $\Kstaticprime(\Vzero)$.
Note that $\tone^{\mathrm{AP2}}\approx\tone^{\mathrm{Vinet}}+1$ whenever the Vinet and AP2 fits yield close values of $\Kprimezero$.
For comparison with the conventional constant-$t$ relations, the effective equilibrium index is defined as
\begin{equation}
\teffzero\equiv t(\Vzero)-3s,
\label{eq:teffzero}
\end{equation}
which gives $\gammazero=\Kprimezero/2-1/6-\teffzero/3.$

The static energy--volume data are fitted independently to the Vinet \cite{vinet1987universal} and AP2 \cite{Holzapfel1998} forms.
The Vinet energy expression is given by
\begin{equation}
\Estatic(V) = \Ezero+\frac{9\Kzero\Vzero}{\eta^2} \left[ 1+\left(\eta(1-x)-1\right) e^{\eta(1-x)}\right]
\label{eq:Vinet}
\end{equation}
with $x \equiv \left(V/\Vzero\right)^{1/3}$ and $\eta \equiv \frac{3}{2}\left(\Kprimezero-1\right)$.
The AP2 energy can be expressed as
\begin{align}
\label{eq:AP2}
\Estatic(V)
&=
\Ezero
-
3\Vzero
\int_{1}^{x}
\Pstatic(u)\,u^2\,du ,
\\
\Pstatic(x)
&=
3\Kzero x^{-5}(1-x)
\exp\!\left[c_0(1-x)\right]
\left[
1+c_2x(1-x)
\right],
\nonumber
\end{align}
with $c_0\equiv\ln\!\left(\frac{\PFGzero}{3\Kzero}\right)$, $c_2\equiv \frac{3}{2}(\Kprimezero-3)-c_0$, and $x$ as defined above.
Here $\PFGzero$ is the nonrelativistic free-electron
Fermi-gas pressure evaluated at $\Vzero$ using the total electron count
in the cell.
The construction of $\gamma(V)$ is carried out separately for the
Vinet and AP2 static baselines, using the corresponding fitted values of
$\Vzero$, $\Kzero$, and $\Kprimezero$ in Eq.~\eqref{eq:tone}.
The fitted static EOS parameters are reported in Table~S1 of the Supplemental Material \cite{supmat}.

The Gr\"uneisen parameter at the static equilibrium volume is defined as
\begin{equation}
\gammazero
=
-
\left.
\frac{\partial \ln \thetaD(V)}{\partial \ln V}
\right|_{V=\Vzero},
\label{eq:gamma0_def}
\end{equation}
and can therefore be obtained from the volume dependence of the Debye temperature
$\thetaD(V)$.
In practice, $\thetaD(V)$ is evaluated at five volumes: 
$\Vzero$, together with two slightly compressed and two slightly expanded volumes on either side of $\Vzero$.
$\gammazero$ is then extracted from a constrained quadratic fit of the form
\begin{equation}
\ln\!\left(\frac{\thetaD(V)}{\thetaDzero}\right)
=
-\gammazero \ln\!\left(\frac{V}{\Vzero}\right)
+
\gtwo \left[\ln\!\left(\frac{V}{\Vzero}\right)\right]^2,
\label{eq:gamma0_fit}
\end{equation}
where $\thetaDzero = \thetaD(\Vzero)$ and
$\gtwo$ accounts for the leading curvature in
$\ln\thetaD$ about $\Vzero$.
Constrained linear fits obtained by setting $\gtwo=0$ yield virtually
the same $\gammazero$ values for all materials and MLIPs studied here,
confirming that the five-volume sampling interval is sufficiently
narrow for a robust extraction of $\gammazero$.

The Debye temperature at each volume is computed from the average sound velocity
$\bar{c}(V)$ according to \cite{grimvall1999thermophysical}
\begin{align}
\theta_D(V)
&=
\frac{\hbar}{\kB}
\left(\frac{6\pi^2 N}{V}\right)^{1/3}
\bar{c}(V),
\label{eq:thetaDV} \\
\bar{c}(V)
&=\left[\frac{1}{3}\sum_{k=1}^{3}\int \frac{1}{[c_k(\mathbf{n})]^3}\,\frac{d\Omega}{4\pi}\right]^{-1/3},
\label{eq:cbar}
\end{align}
where $\hbar$ is the reduced Planck constant, $\kB$ is Boltzmann's constant, and
$N$ is the number of atoms in the cell of volume $V$.
The three acoustic phase velocities $\{c_k(\mathbf{n})\}_{k=1}^{3}$ for propagation direction
$\mathbf{n}$ are obtained by solving the Christoffel equation \cite{Musgrave1970}:
\begin{equation}
\left(\Gamma_{ik}(\mathbf{n})-\rho\,c^2 \delta_{ik}\right) u_k = 0,
\label{eq:chris}
\end{equation}
where $\Gamma_{ik}(\mathbf{n}) = C_{ijkl}\, n_j n_l$,
$C_{ijkl}$ is the elastic stiffness tensor,
$\rho$ is the mass density, and
$u_k$ is the polarization eigenvector.
The integration over solid angle $\Omega$ in Eq.~\eqref{eq:cbar} is evaluated in the present work
by Monte Carlo sampling of 20\,000 uniformly distributed propagation directions $\mathbf{n}$ on the unit sphere.
Increasing the angular grid from 20\,000 to 32\,000 directions changed the Debye temperatures by less than $0.02$~K for all materials studied, confirming that the 20\,000-direction sampling is sufficiently converged.
Although full elastic anisotropy is retained through the direction-dependent Christoffel velocities, its reduction to a single average acoustic scale in the Debye description may become less accurate for lower-symmetry or strongly anisotropic materials.

The elastic stiffness tensor is evaluated separately for each selected
volume in the near-equilibrium range.
At volumes away from $\Vzero$, the quantity required for constructing
$\thetaD(V)$ is the strain--energy curvature about a structure at a
prescribed volume, rather than the thermodynamic elastic response under
an applied external pressure.
Accordingly, the fits use total energies rather than enthalpies; adding
a $PV$ term would define a different finite-pressure response.
This treatment parallels quasiharmonic phonon calculations, in which
force constants and vibrational frequencies are evaluated at imposed
volumes.

Starting from the internally relaxed reference structure at a given
fixed volume, small strain amplitudes are applied according to the
irreducible strain modes of the corresponding Laue class, as obtained
from character tables \cite{Aroyo2006Bilbao}.
The deformation modes are used in the form provided by the symmetry
analysis, i.e., without normalizing the relative magnitudes of their
strain components; the overall deformation amplitude is controlled by
a scalar strain parameter.
For each imposed deformation amplitude, internal atomic coordinates are
relaxed and the total energy $E(\boldsymbol{\varepsilon})$ is evaluated.
The elastic constants are determined by fitting the quadratic
elastic-energy expression
\begin{equation}
E(\boldsymbol{\varepsilon})
=
E_0
+
\tfrac{1}{2} V\,
\boldsymbol{\varepsilon}^{\mathrm T}
C
\boldsymbol{\varepsilon},
\label{eq:eefit}
\end{equation}
to the calculated energies using a weighted fitting procedure, in which
configurations with smaller energy increments relative to the unstrained
reference are assigned larger weight.
The fit is performed under symmetry constraints appropriate to the
Laue class \cite{Mouhat2014Necessary}, yielding a stiffness tensor $C$
consistent with the crystal symmetry.

Before fitting, contributions odd in strain are removed by averaging
the energies of each complete opposite-strain pair.
For each nonzero pair, the fitting target is
\begin{equation}
E_i
=
\frac{
E(+\boldsymbol{\varepsilon}_i)
+
E(-\boldsymbol{\varepsilon}_i)
}{2},
\end{equation}
while the unstrained configuration is included as a separate target.
The independent components of $C$ and the fitted intercept $E_0$ are
then determined simultaneously.
The weighted mixed-norm objective $\Phi$ is minimized using the
Nelder--Mead simplex algorithm~\cite{Nelder1965}, where
\begin{equation}
\begin{aligned}
\Phi
&=
\lambda \sum_i w_i |r_i|
+
(1-\lambda)
\left(\sum_i w_i r_i^2\right)^{1/2},
\\
w_i
&=
\frac{\exp(-\beta\Delta E_i)}
     {\sum_j \exp(-\beta\Delta E_j)}.
\end{aligned}
\label{eq:weighted_elastic_fit}
\end{equation}
Here,
\begin{equation}
r_i=E_i^{\mathrm{fit}}-E_i,
\qquad
\Delta E_i=E_i-E_{\mathrm{ref}}(V),
\end{equation}
where $\beta$ is an inverse-energy decay parameter and
$E_{\mathrm{ref}}(V)$ is the energy of the unstrained, internally
relaxed reference structure at the same imposed volume.
In all fits, $\beta=40~\mathrm{eV}^{-1}$ and $\lambda=0.5$ are used.
The calculated strain--energy data, opposite-strain-symmetrized fitting
targets, and resulting weighted quadratic fits at all five sampled
volumes are shown for each material--potential combination in
Figs.~S1--S12 of the Supplemental Material~\cite{supmat}.

In summary, the workflow for constructing the Gr\"uneisen function is as follows:
\begin{enumerate}
  \item Fully relax the structure (cell and ionic degrees of freedom) to obtain the equilibrium configuration.
  \item Generate the static energy--volume data and fit them separately to the Vinet and AP2 EOS forms [Eqs.~\eqref{eq:Vinet} and \eqref{eq:AP2}, respectively].
  \item For five volumes in the vicinity of the equilibrium volume:
  \begin{enumerate}
    \item Obtain the elastic stiffness tensor by fitting the strain--energy data [Eq.~\eqref{eq:eefit}].
    \item Solve the Christoffel equation [Eq.~\eqref{eq:chris}] to obtain the three acoustic phase velocities.
    \item Evaluate the average sound velocity from Eq.~\eqref{eq:cbar}.
    \item Compute the Debye temperature using Eq.~\eqref{eq:thetaDV}.
  \end{enumerate}
  \item Extract $\gammazero$ from the constrained fit in Eq.~\eqref{eq:gamma0_fit}.
  \item Compute $\tone$ from Eq.~\eqref{eq:tone}.
  \item Evaluate $t(V)$ from Eq.~\eqref{eq:tV} with $\tzero=5/2$.
  \item Evaluate $\gammaeos(V)$ using Eq.~\eqref{eq:gamma_eos}.
  \item Construct $\gamma(V)$ from Eq.~\eqref{eq:gamma}, using $s=1/3$ for the Vinet EOS and $s=0$ for the AP2 EOS.
\end{enumerate}

\subsection{Calculation of finite-temperature bulk moduli}

A Mie--Gr\"uneisen EOS is employed with a Debye thermal term \cite{ZharkovKalinin1971}
\begin{equation}
P(V,T) = \Pstatic(V) + \gamma(V)\frac{\EDebye(V,T)}{V}, 
\label{eq:miegru}
\end{equation}
where 
\begin{equation}
\EDebye(V,T) = N \kB \left[ \frac{9}{8}\,\thetaD + 3T\,\Debye\!\left(\frac{\thetaD}{T}\right) \right]
\end{equation}
is the Debye vibrational internal energy (including the zero-point term), with $\Debye(x)$ denoting the Debye function of order three.

The Debye temperature is evaluated using
\begin{equation}
\thetaD(V)
= \thetaDzero\,\left( \frac{V}{\Vzero} \right)^{\frac{1}{6}-s}
\left[
\frac{\Kstatic - \frac{2}{3}\,t\,\Pstatic}{\Kzero}
\right]^{\frac{1}{2}},
\label{eq:thetaD_final}
\end{equation}
which follows by substituting Eq.~\eqref{eq:gamma} into
\begin{equation}
\thetaD(V)= \thetaD(\Vzero) \,\exp\!\left[-\int_{\Vzero}^{V}\gamma(V')\,\frac{dV'}{V'}\right].
\end{equation}

In this framework, the adiabatic and isothermal bulk moduli are given by
\begin{align}
\Kadiabatic(V,T)&=\Kstatic(V) + \frac{\gamma}{V}\bigl(1-q+\gamma\bigr)\EDebye \label{eq:KS}\\
\Kisothermal(V,T)&=\Kadiabatic(V,T)-\frac{\gamma^2\,\CV}{V}\,T,\label{eq:KT}
\end{align}
respectively.
Here
$q(V) = d\ln \gamma / d\ln V$
and
\begin{equation}
\CV(V,T)
= 3 N \kB\left[
4\,\Debye \left(\frac{\thetaD}{T}\right)
- \frac{3\,(\thetaD/T)}{e^{\thetaD/T}-1}
\right]
\end{equation}
is the heat capacity at constant volume \cite{anderson1995equations}.

To obtain the temperature dependence of the bulk moduli at zero pressure,
the equilibrium volume $\Ve(T)$ is determined by solving $P(\Ve,T)=0$ over a range of temperatures.
Evaluating Eqs.~\eqref{eq:KS} and \eqref{eq:KT} at $V=\Ve(T)$ then yields
$\Kadiabatic(T)\equiv \Kadiabatic(\Ve(T),T)$ and
$\Kisothermal(T)\equiv \Kisothermal(\Ve(T),T)$, respectively.
Note that $\Kadiabatic(0) = \Kisothermal(0)$
but $\Kadiabatic(0) \ne \Kzero$ because the zero-point vibrations shift the zero-pressure equilibrium volume, i.e., $\Ve(0) \ne \Vzero$.

For compact representation of the calculated bulk-modulus curves and for obtaining smooth temperature derivatives, the computed $\Kadiabatic(T)$ and $\Kisothermal(T)$ values are additionally parameterized by
\begin{equation}
K(T)=K(0)- b\,T\,e^{-\Tzero/T},
\label{eq:Wachtman}
\end{equation}
with $b$ and $\Tzero$ obtained from separate fits to the calculated curves in each case.
This auxiliary parametrization is used only for interpolation and analytic evaluation of temperature derivatives; it does not enter the construction of $\gamma(V)$ or the Mie--Gr\"uneisen--Debye prediction.
This form, introduced by Wachtman \textit{et al.} \cite{Wachtman1961} to fit the temperature dependence of Young's modulus in oxides,
was later theoretically justified for the bulk modulus \cite{anderson1966}
and extended to other materials \cite{Nandanpawar1978} and other elastic constants \cite{grimvall1999thermophysical}.
From the fitted values of $b$ and $\Tzero$, the temperature derivative is evaluated analytically as
$
dK/dT = -b\,e^{-\Tzero/T}\left(1+\Tzero/T\right).
$
At high temperature ($T \gg \Tzero$), Eq.~\eqref{eq:Wachtman} can be linearized as
$
K(T) \approx K(0) - b\,(T-\Tzero),
$
so that the parameter $b$ sets the high-temperature slope, $dK/dT \approx -b$.

The thermal softening rate $dK/dT$ is experimentally accessible and is thus well suited for comparison with predictions.
Because $dK/dT$ emphasizes the temperature dependence rather than the absolute magnitude, it provides a direct test of the temperature-dependent part of the present approach and is largely insensitive to any systematic offset in $K$.

\subsection{\label{sec:mlip}Machine-learning interatomic potential calculations}

\begin{figure}[b]
\includegraphics[width=0.47\textwidth]{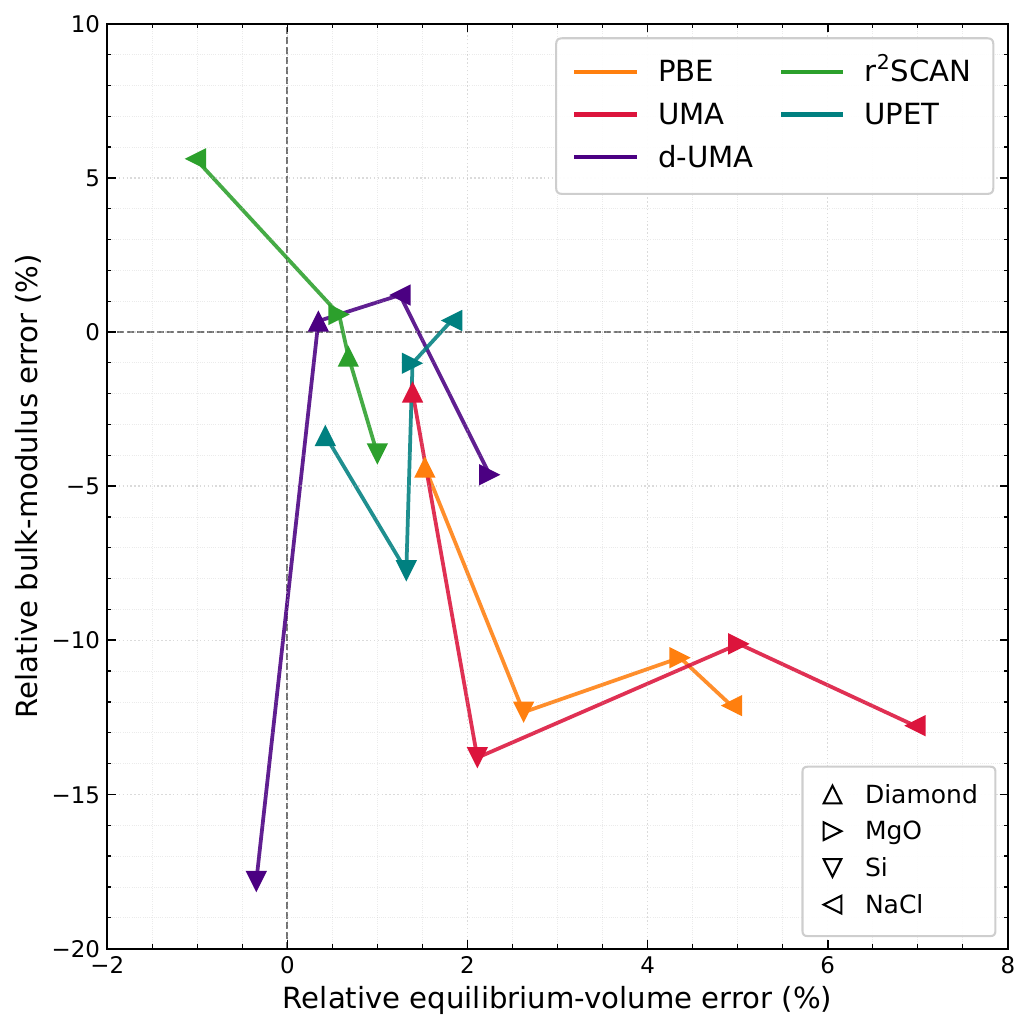}
\caption{
Relative errors in the static bulk modulus,
$(\Kzerocalc/\Kzeroexpt-1)\times 100$,
versus equilibrium volume,
$(\Vzerocalc/\Vzeroexpt-1)\times 100$,
for diamond, MgO, Si, and NaCl.
PBE, r$^2$SCAN, and the zero-point-corrected experimental reference values
are taken from Ref.~\onlinecite{kaplan2022laplacian}.
The connecting lines are guides to the eye.
}
\label{fig:static_errors}
\end{figure}

MLIP-based atomistic calculations were carried out using the Atomic Simulation Environment
(ASE; v3.26.0) \cite{larsen2017atomic} as the workflow engine.
For comparison, energies, atomic forces, and stress tensors were evaluated using two universal
MLIP frameworks. First, calculations were performed with the FAIRChem library
(\texttt{fairchem-core}, v2.12.0) \cite{fairchem}, through its ASE interface
(\texttt{FAIRChemCalculator}), employing the Universal Models for Atoms (UMA) with the pretrained
checkpoint \texttt{uma-s-1p1} \cite{wood2025uma}. UMA was selected in part because its architecture
builds on the equivariant Smooth Energy Network (eSEN) \cite{fu2025esen}, whose smooth energy
landscape is advantageous for evaluating derivative-based properties. 
Second, calculations were performed with the UPET package (\texttt{upet}, v0.2.5), through its ASE-compatible \texttt{UPETCalculator}, 
employing the PET-MAD-1.5 (v1.5.0) model with the pretrained \texttt{pet-mad-s} checkpoint \cite{PET-MAD-1.5-2026}, implemented within the current UPET architecture \cite{pushing-unconstrained-2026}.
The two MLIPs differ not only in architecture and training strategy, but also in the underlying reference data:
UMA is trained on reference data generated with the Perdew--Burke--Ernzerhof (PBE) generalized-gradient-approximation (GGA) functional \cite{perdew1996generalized}, 
whereas PET-MAD-1.5 is trained on reference data generated the regularized-restored strongly constrained and appropriately normed (r$^2$SCAN) meta-GGA functional \cite{furness2020r2scan}.

Figure~\ref{fig:static_errors} shows that UMA closely follows the qualitative PBE error pattern for the present set of solids: 
the equilibrium volumes are systematically overestimated and the corresponding static bulk moduli are underestimated.
This behavior is consistent with UMA providing a PBE-level static-EOS baseline.
UPET substantially reduces the errors, placing most points closer to the origin in Fig.~\ref{fig:static_errors}.
The UPET points do not, however, simply reproduce the material-by-material r$^2$SCAN trend, 
indicating that the improvement is not merely a trivial transfer of the tabulated r$^2$SCAN behavior 
but reflects the independent accuracy and residual biases of the PET-MAD-1.5 model.

\begin{figure}
\includegraphics[width=0.47\textwidth]{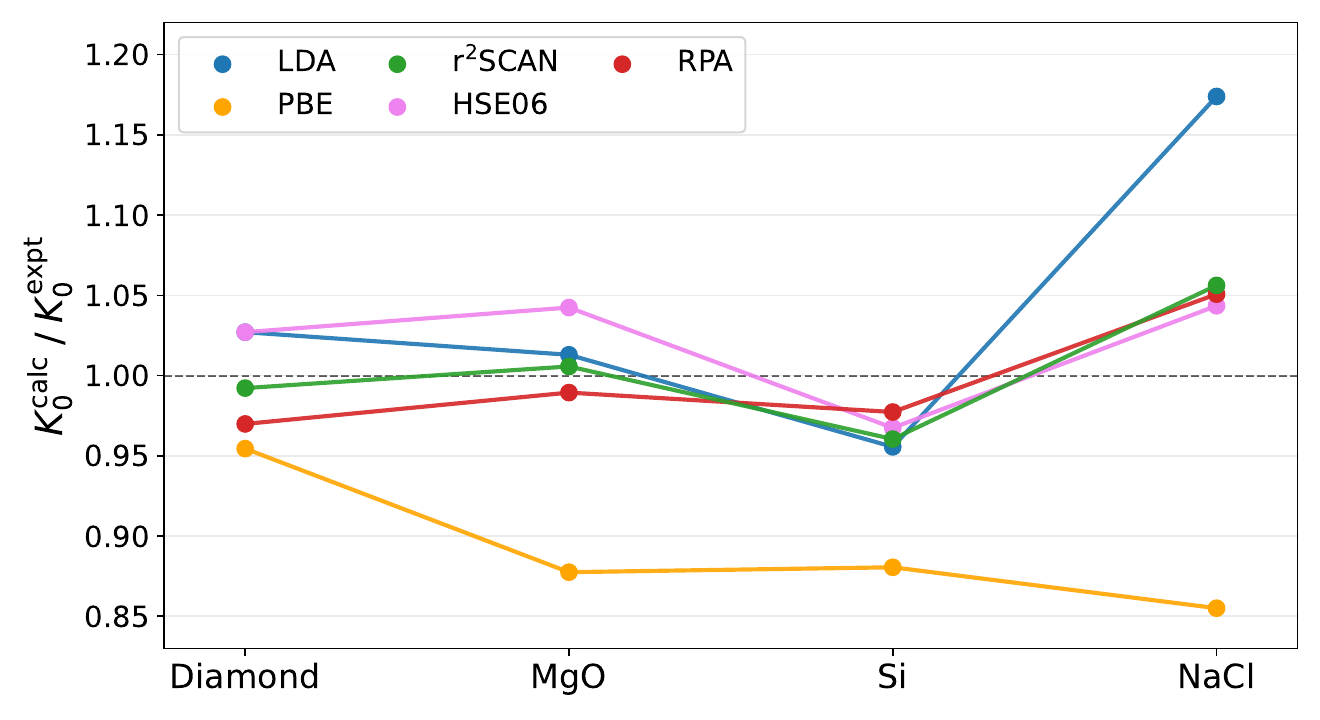}
\caption{
Ratio $\Kzerocalc/\Kzeroexpt$ for diamond, MgO, Si, and NaCl
obtained with several electronic-structure approximations.
The calculated static bulk moduli $\Kzerocalc$ are taken from
Ref.~\onlinecite{Csonka2009Assessing} for LDA and PBE,
Ref.~\onlinecite{kaplan2022laplacian} for r$^2$SCAN,
Ref.~\onlinecite{schlipf2011hse} for HSE06, and
Ref.~\onlinecite{harl2010assessing} for RPA.
The zero-point-corrected experimental reference values $\Kzeroexpt$
are taken from Ref.~\onlinecite{tran2016rungs}.
The connecting lines are guides to the eye.
}
\label{fig:K_comp}
\end{figure}

Figure~\ref{fig:K_comp} shows that PBE systematically underestimates $\Kzero$ for all solids considered here.
This underestimation follows a distinct trend that is not shared by the simpler local-density approximation (LDA), nor by more sophisticated approaches, including the meta-GGA r$^2$SCAN functional, the range-separated HSE06 hybrid functional, and the random-phase approximation (RPA).
Such systematic PBE-level EOS errors are commonly treated by using correction schemes or by applying pressure or volume shifts to the calculated EOS, as in Refs.~\onlinecite{Oganov2003} and \onlinecite{Wu2008Wentzcovitch}.

To partly mitigate the PBE-inherited EOS bias in UMA, a dispersion-corrected variant (d-UMA) was also considered, motivated by earlier studies based on dispersion-corrected DFT and MLIPs \cite{gulerkilic2015,kilic2024}.
For the d-UMA calculations, \texttt{FAIRChemCalculator} was combined with the ASE calculator provided by the Simple DFT-D3 library (\texttt{simple-dftd3}, v1.2.1) \cite{Ehlert2024JOSS}, using the DFT-D3(BJ) correction with PBE damping parameters \cite{grimme2011effect}.
Because long-range dispersion interactions are not explicitly represented in UMA \cite{wood2025uma}, the D3 correction was applied to assess their effect, particularly for the more ionic solids MgO and NaCl.
For the covalent materials diamond and Si, the effect of D3 is comparatively subtle and not systematic: although it can shift the equilibrium volume, it does not necessarily improve the EOS curvature.
Nevertheless, for completeness and comparison, results from d-UMA alongside those from UMA and UPET are reported for all solids considered here.
Using UMA, UPET, and d-UMA comparatively enables one to assess the sensitivity of the predicted thermal evolution to the static EOS baseline.

Structure optimizations were performed 
using the accelerated bias-corrected variant of the fast inertial relaxation engine (ABC-FIRE) \cite{echeverri2023abcfire,guenole2020fire2,bitzek2006fire}, 
as implemented in the \texttt{FIRE2} class of ASE.
The \texttt{FIRE2} constructor parameters were set to 
\texttt{dt}=0.03, \texttt{maxstep}=0.02, \texttt{dtmax}=0.18, \texttt{dtmin}=$10^{-3}$, \texttt{Nmin}=35, \texttt{finc}=1.05, \texttt{fdec}=0.75, \texttt{astart}=0.12, \texttt{fa}=0.98, and 
\texttt{use\_abc}=\texttt{true}; these settings were found to yield stable convergence across the optimization runs.
Because ASE employs a single generalized-force convergence parameter (\texttt{fmax}), the ABC-FIRE optimizations were performed iteratively, monitoring both the maximum atomic force ($f_{\max}$) and the maximum absolute component of the unit-cell stress tensor ($\sigma_{\max}$).
The thresholds for $f_{\max}$ and $\sigma_{\max}$ were set to
$5\times10^{-4}$~eV/\AA{} and $5\times10^{-3}$~GPa, respectively.
First, crystal structures were fully relaxed, allowing both the unit-cell vectors and ionic positions to vary.
Next, the static energy--volume data used in the Vinet and AP2 fits
[Eqs.~\eqref{eq:Vinet} and \eqref{eq:AP2}] were generated from
fixed-volume optimizations in which the cell volume was held constant
while the cell shape and internal coordinates were relaxed subject to
a symmetry-preserving constraint enforced using the
\texttt{FrechetCellFilter} class of ASE.
Finally, the energy--strain data required for the elastic energy fitting [Eq.~\eqref{eq:eefit}] were generated by applying symmetry-adapted strains to the optimized structures at each of the five volumes used to extract $\gamma_0$, with internal coordinates relaxed at each strain.
The EOS dataset was sampled over a broad range of volume scalings around equilibrium, while excluding the most extreme
compression and dilation regimes where the MLIP may be extrapolating beyond its reliable domain; elastic strains were
restricted to the small-deformation regime to remain in the linear elastic response and to avoid strain-induced mechanical instabilities.

\section{\label{s:res}RESULTS AND DISCUSSION}

In this section, the results obtained from the present construction of $\gamma(V)$ are presented and discussed.
The finite-temperature bulk moduli and their thermal softening are considered first, since they provide a direct validation of the present framework against experiment and allow comparison with available finite-temperature computational results.
The corresponding behavior of $\gamma$, $q$, and $\thetaD$ is then analyzed to elucidate the physical character of the resulting Gr\"uneisen functions.
This order of presentation is adopted because experimental estimates of $\gamma$, $q$, and $\thetaD$ are often scattered, as these quantities are not directly measurable and must instead be inferred from other data in a model-dependent manner, with their extraction also depending on how temperature effects are treated.

\begin{figure*}
  \includegraphics[width=0.93\textwidth]{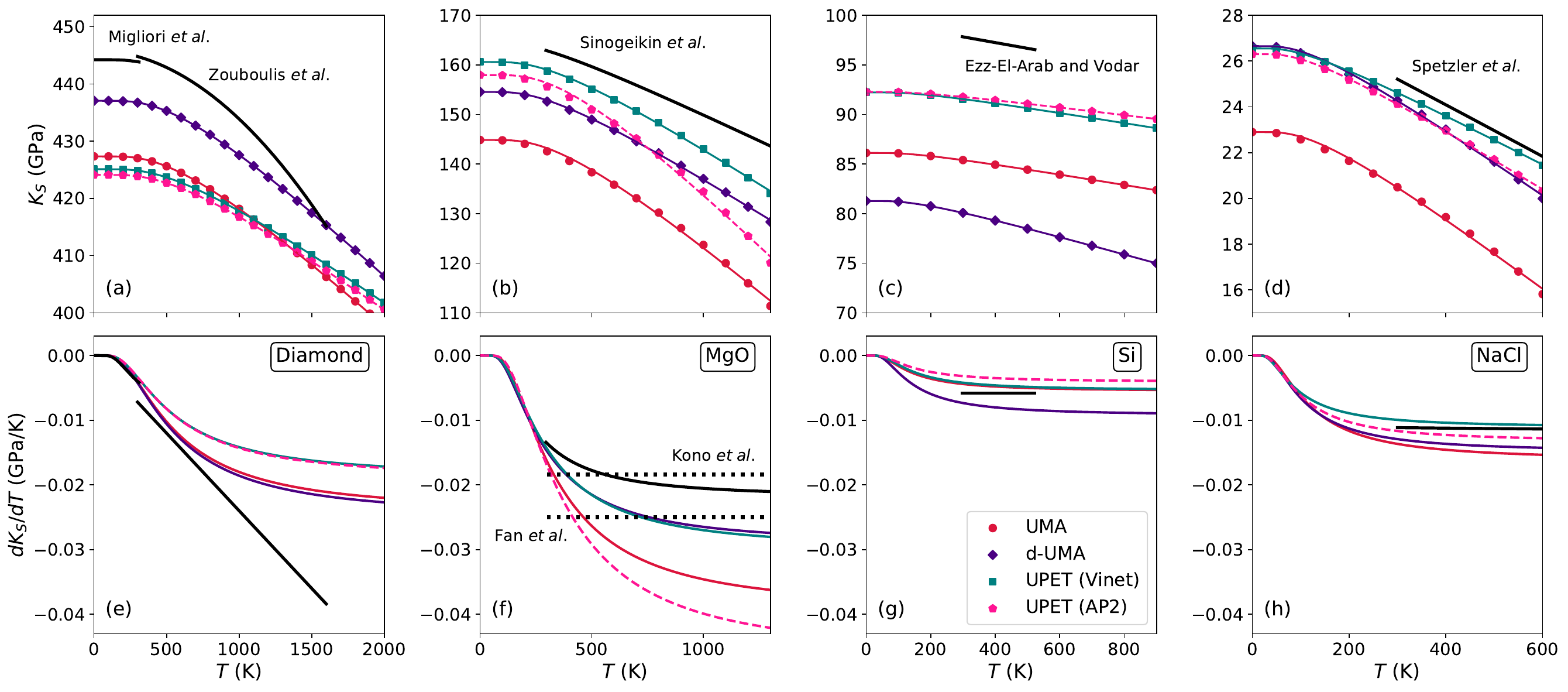}
\caption{
Temperature dependence of the adiabatic bulk modulus $\Kadiabatic(T)$
[panels (a)--(d), top row] and its temperature derivative
$d\Kadiabatic/dT$ [panels (e)--(h), bottom row]
for diamond, MgO, Si, and NaCl, obtained within the
Mie--Gr\"uneisen--Debye framework using the present $\gamma(V)$
construction.
Filled symbols denote results obtained using the UMA,
dispersion-corrected UMA (d-UMA), and UPET potentials.
For UPET, results obtained using the Vinet and AP2 equations of state
are shown.
Curves passing through the symbols are fits to
Eq.~\eqref{eq:Wachtman}; the UPET(AP2) fits are shown as dashed curves,
whereas the other calculated fits are solid.
The derivatives are obtained analytically from these fits.
Experimental curves are reproduced using data or fitting results from
Migliori \textit{et al.} (Ref.~\onlinecite{Migliori2008}),
Zouboulis \textit{et al.} (Ref.~\onlinecite{Zouboulis1998}),
Sinogeikin \textit{et al.} (Ref.~\onlinecite{Sinogeikin2000Compact}),
Ezz-El-Arab and Vodar (as given in Ref.~\onlinecite{Nandanpawar1978}), and
Spetzler \textit{et al.} (Ref.~\onlinecite{Spetzler1972Equation}).
The black curves and line segments in the bottom row are analytic
derivatives of the corresponding experimental curves in the top row.
The horizontal dotted lines in panel (f) mark the experimental
$d\Kadiabatic/dT$ values reported by
Kono \textit{et al.} (Ref.~\onlinecite{Kono2010PETI}) and
Fan \textit{et al.} (Ref.~\onlinecite{Fan2019AmMin}).
}
  \label{fig:K_vs_T}
\end{figure*}

Figure~\ref{fig:K_vs_T} shows the adiabatic bulk modulus
$\Kadiabatic(T)$ (top row) and its temperature derivative
$d\Kadiabatic/dT$ (bottom row) for diamond, MgO, Si, and NaCl,
computed using the UMA, d-UMA, and UPET potentials.
For UPET, results obtained using both the Vinet and AP2 equations of
state are included to assess the sensitivity of the thermal predictions
to the analytical EOS entering the $\gamma(V)$ construction.
The corresponding fit parameters from Eq.~\eqref{eq:Wachtman},
including both adiabatic and isothermal cases, are listed in
Table~S2 \cite{supmat}.
Experimental curves are included for comparison using the
parametrizations given in Table~S3 \cite{supmat}.
For all four solids, the predictions reproduce the observed thermal-softening trends and yield $d\Kadiabatic/dT$ values in reasonable quantitative agreement with experiment at low and moderate temperatures, although the degree of agreement necessarily depends on the material, the underlying potential, and the choice of analytical EOS.

For diamond, all calculations reproduce the expected monotonic thermal
softening, while d-UMA shifts the absolute $\Kadiabatic(T)$ curve closest
to the experimental parametrizations.
The calculated $d\Kadiabatic/dT$ nevertheless remains less negative than
the experimental derivative at high temperature, particularly for the
UPET results; the UPET(Vinet) and UPET(AP2) curves are nearly
indistinguishable.
Since the high-temperature variation of diamond's bulk modulus cannot be
fully captured within a simple Mie--Gr\"uneisen--Debye framework
\cite{Dewaele2008High}, the discrepancy is quantified near the
characteristic temperature scale of the fitted curves.
The fitted $\Tzero$ values are $965$~K for UMA, $963$~K for d-UMA,
$937$~K for UPET(Vinet), and $947$~K for UPET(AP2).
When evaluated at their respective $\Tzero$ values, UMA, d-UMA,
UPET(Vinet), and UPET(AP2) reproduce approximately $76\%$, $79\%$,
$61\%$, and $61\%$, respectively, of the experimental softening
magnitude derived from Ref.~\onlinecite{Zouboulis1998}.
The remaining discrepancy can be attributed to contributions that are
omitted or imperfectly represented within the present quasiharmonic-level
description, most notably intrinsic anharmonic effects.
Nevertheless, despite neglecting explicit anharmonicity, the present
approach captures a substantial fraction of the thermal softening of
$\Kadiabatic(T)$ up to temperatures near $\Tzero$.

For MgO, the improvement obtained with d-UMA relative to UMA is again
substantial: the offset in $\Kadiabatic(T)$ is reduced, and the predicted
$d\Kadiabatic/dT$ lies much closer to the experimentally inferred range.
UPET(Vinet) further improves the absolute bulk modulus and gives a
thermal-softening rate comparable to that obtained with d-UMA.
UPET(AP2), however, predicts substantially stronger thermal softening,
and its $\Kadiabatic(T)$ curve departs increasingly from experiment as
the temperature rises.

For silicon, the calculated temperature dependence remains physically
reasonable in all cases.
The absolute modulus is significantly underestimated by UMA and d-UMA,
whereas both UPET results reduce this offset.
UPET(Vinet) gives a thermal-softening rate close to the experimental
slope, while UPET(AP2) predicts somewhat weaker softening.

For NaCl, d-UMA markedly improves the absolute bulk modulus relative to
UMA, although both potentials predict stronger thermal softening than
indicated by experiment.
The two UPET results further improve the absolute $\Kadiabatic(T)$ values
and are comparatively close to one another, with UPET(AP2) predicting
slightly stronger softening than UPET(Vinet).

Collectively, these results indicate that the present $\gamma(V)$
construction captures the overall temperature dependence of
$\Kadiabatic$ across chemically and mechanically distinct solids.
The residual offsets in the absolute bulk modulus remain largely
attributable to the static description supplied by the underlying
potential.
The comparison between UPET(Vinet) and UPET(AP2) further shows that the
predicted thermal-softening rate can depend on the selected EOS, most
noticeably for MgO and more weakly for the other materials.
Thus, eliminating the asymptotic shift required by the Vinet
construction does not by itself produce systematically better
finite-temperature agreement.

\begin{figure}
\includegraphics[width=0.47\textwidth]{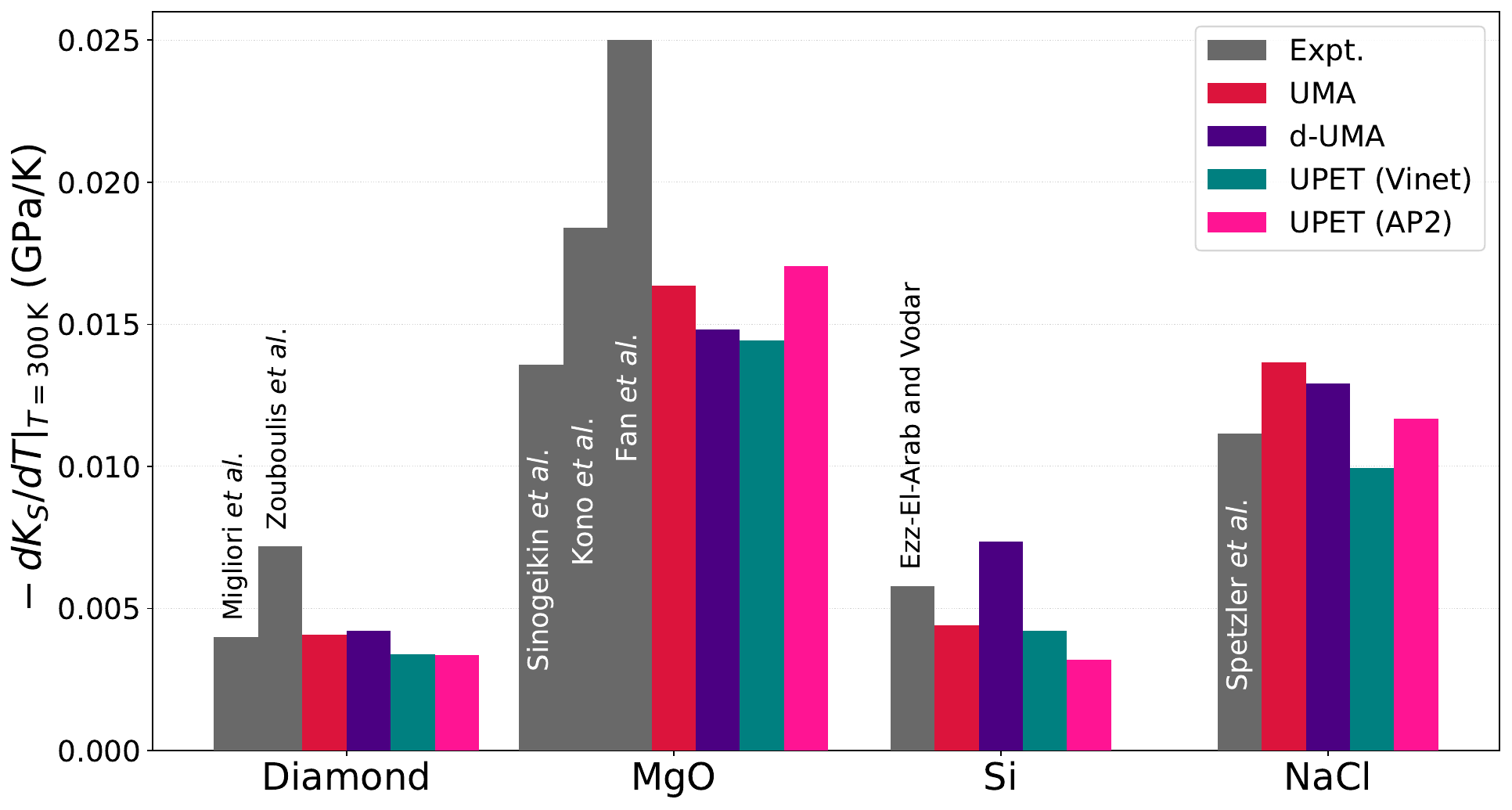}
\caption{
Room-temperature thermal-softening rate of the adiabatic bulk modulus,
$\left.-d\Kadiabatic/dT\right|_{T=300\;\mathrm{K}}$.
Values are derived from the results shown in
Fig.~\ref{fig:K_vs_T}(e)--(h).
}
\label{fig:dK_S_dT}
\end{figure}

Figure~\ref{fig:dK_S_dT} summarizes the room-temperature magnitude of
the adiabatic thermal-softening rate,
$\left.-d\Kadiabatic/dT\right|_{T=300\;\mathrm{K}}$,
obtained from the calculated and experimental results presented in
Fig.~\ref{fig:K_vs_T}.
Across the four systems, the calculated values are generally in
reasonable quantitative agreement with the available experimental
estimates and reproduce the experimental relative ordering across the
four solids.
For diamond, the UMA and d-UMA values lie close to the lower of the two
experimental estimates, whereas both UPET variants predict slightly
weaker softening and yield nearly identical values.
For MgO, the experimental determinations exhibit a comparatively broad
spread, and all calculated values fall within the range spanned by the
experimental estimates.
Notably, for both diamond and MgO, the spread among the calculated
values is smaller than that among the available experimental
determinations.
For Si and NaCl, conversely, the experimental estimates fall within the
range spanned by the calculated values.

As discussed above, the degree of quantitative agreement depends not only
on the material and the underlying potential but also on the choice of
analytical EOS.
Nevertheless, at room temperature, the comparison between the two UPET
variants reveals no systematic improvement upon replacing the Vinet EOS
with AP2.
Relative to UPET(Vinet), UPET(AP2) produces almost no change for diamond,
reduces the predicted thermal-softening rate for Si, increases it
appreciably for MgO, and yields a more modest increase for NaCl.

\begin{figure}
\includegraphics[width=0.47\textwidth]{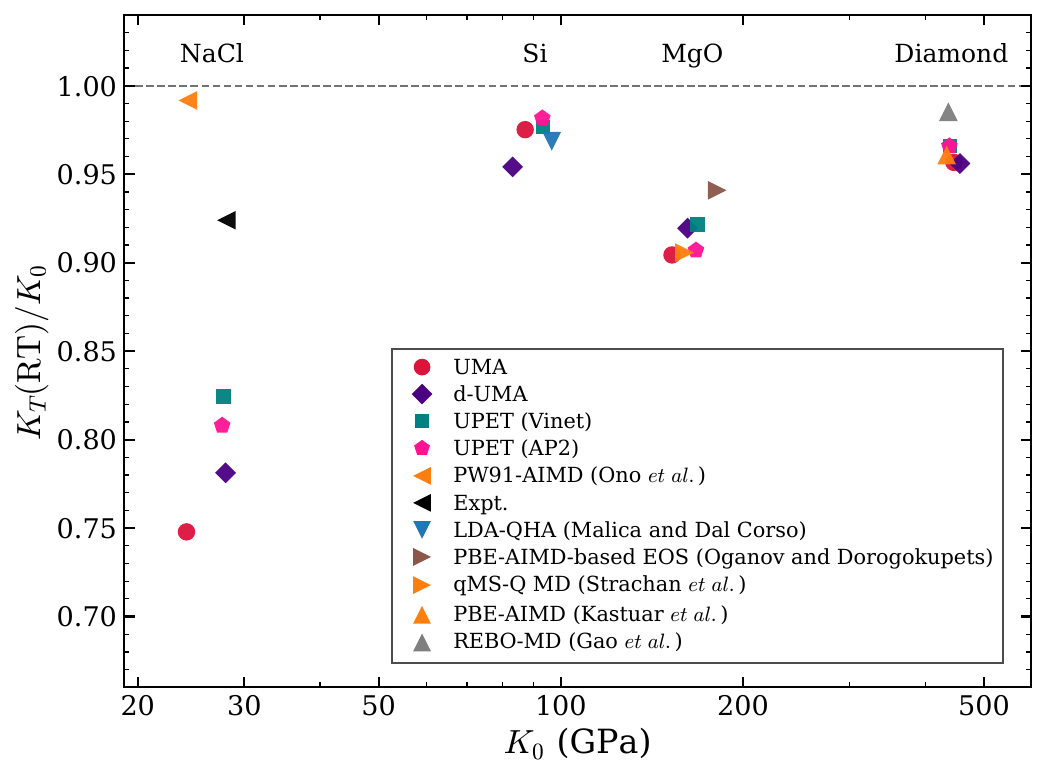}
\caption{
The ratios of room-temperature (RT) isothermal to static bulk moduli,
$\Kisothermal(\mathrm{RT})/K_0$, for NaCl, Si, MgO, and diamond.
The present results are compared with values extracted from other
finite-temperature computational studies:
PW91-AIMD for NaCl (Ref.~\onlinecite{ono2008}),
LDA-QHA for Si (Ref.~\onlinecite{malica2020quasi}),
PBE-AIMD-based EOS for MgO (Ref.~\onlinecite{Oganov2003}),
qMS-Q MD for MgO (Ref.~\onlinecite{strachan1999phase}),
PBE-AIMD for diamond (Ref.~\onlinecite{Kastuar2023}),
and REBO-MD for diamond (Ref.~\onlinecite{gao2006elastic}).
The experimental NaCl point is obtained from fits to the data of
Ref.~\onlinecite{brown1999}.
The dashed line denotes $\Kisothermal(\mathrm{RT})/K_0=1$.
}
\label{fig:KTvsK0}
\end{figure}

In Fig.~\ref{fig:KTvsK0}, the thermal softening of the isothermal bulk
modulus $\Kisothermal$ is quantified by the room-temperature
(RT)-to-static bulk-modulus ratio,
$\RKRT \equiv \Kisothermal(\mathrm{RT})/\Kzero$, and compared with
values obtained from molecular-dynamics (MD) simulations, both
classical MD and AIMD, and from a first-principles
quasiharmonic-approximation (QHA) calculation.
Using this normalized ratio prevents absolute baseline errors in
$\Kzero$ [see Figs.~\ref{fig:static_errors} and
\ref{fig:K_comp}] from obscuring the comparison of fractional
thermal-softening trends.
A direct finite-temperature MD comparison using the same MLIPs would
provide a useful additional benchmark, but it lies outside the present
scope.
For NaCl, the AIMD results of Ono \textit{et al.}~\cite{ono2008},
obtained with the PW91 GGA functional~\cite{PW91}, give
$\RKRT=0.992$, implying very weak thermal softening.
This value substantially underestimates the softening inferred from
experimental compressibility data~\cite{slater1924} and
finite-temperature phonon thermodynamics~\cite{allen2015}.
The experimental NaCl point in Fig.~\ref{fig:KTvsK0} is based on the
data of Ref.~\onlinecite{brown1999}: interpolation of the zero-Kelvin
compression curve gives $\Kzero=28.02$~GPa, while a BM3 fit to the RT
isotherm gives $\Kisothermal(\mathrm{RT})=25.89$~GPa, resulting in
$\RKRT=0.924$.
All four present values lie below this experimental estimate,
indicating an overestimate of the room-temperature softening.
The two UPET results lie appreciably closer to experiment than the UMA
and d-UMA results, with UPET(Vinet) giving the closer agreement of the
two EOS choices.
For Si, all four present values remain close to the first-principles
LDA-QHA result of Ref.~\onlinecite{malica2020quasi}, despite the spread
in the static bulk-modulus baselines.
For MgO, the present values lie within or very near the range spanned
by the MD-based estimates of Refs.~\onlinecite{Oganov2003} and
\onlinecite{strachan1999phase}.
For diamond, all four ratios lie close to unity, as expected for a
very stiff covalent solid with weak fractional room-temperature
softening.
UMA and d-UMA lie close to the PBE-AIMD estimate of
Ref.~\onlinecite{Kastuar2023}.
The two UPET results are nearly coincident and give slightly larger
ratios, lying between the PBE-AIMD and REBO-MD results but closer to
the former.

The comparison between UPET(Vinet) and UPET(AP2) reveals a
material-dependent but minor EOS sensitivity.
Relative to UPET(Vinet), UPET(AP2) gives slightly stronger
room-temperature softening for NaCl and MgO, slightly weaker softening
for Si, and almost no change for diamond.
The AP2 construction therefore does not produce a systematic
improvement relative to Vinet, although both EOS choices yield
room-temperature softening ratios on the same characteristic scale.

\begin{figure*}
\includegraphics[width=0.93\textwidth]{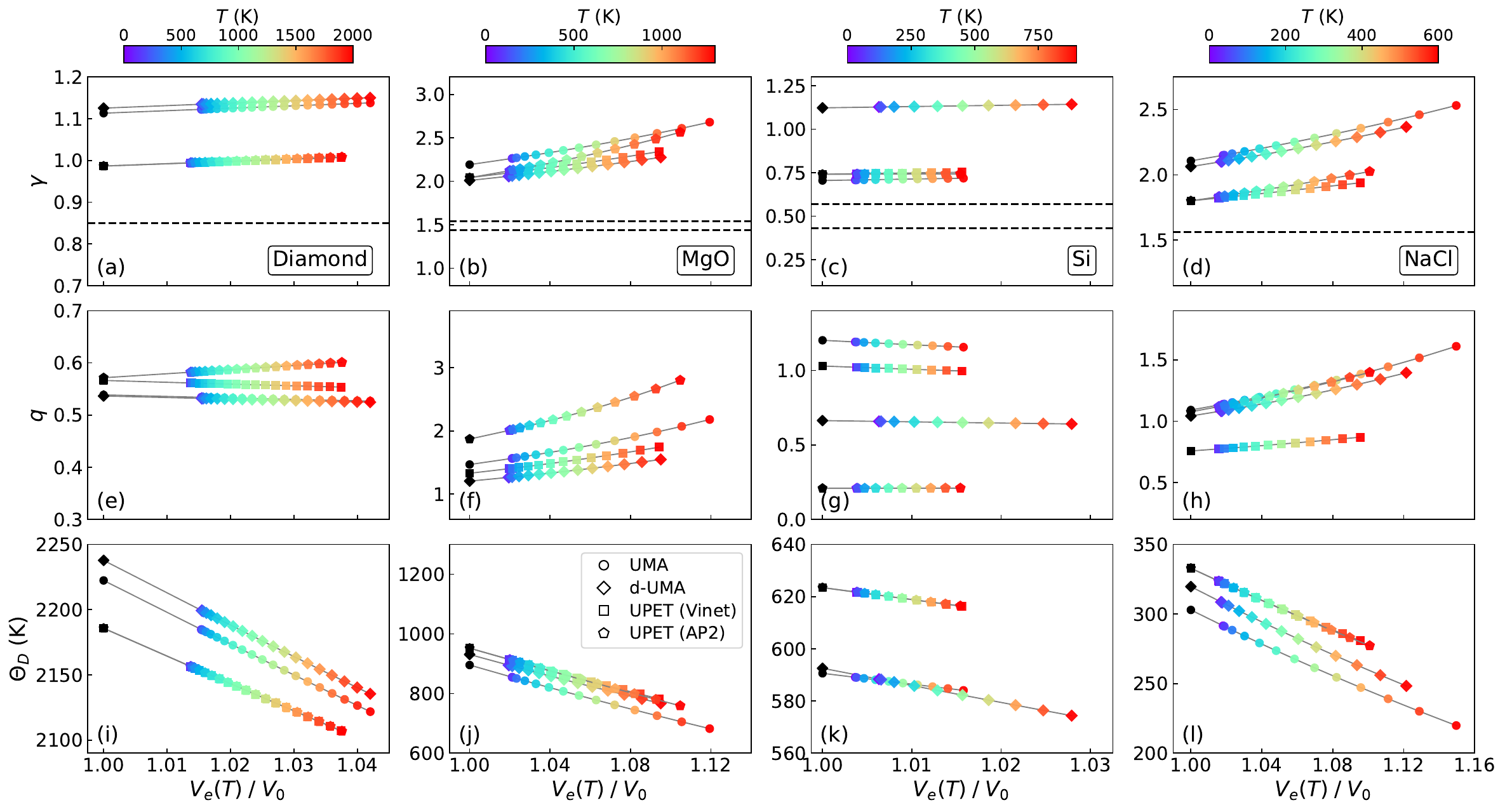}
\caption{
The Gr\"uneisen parameter $\gamma$ [(a)--(d)], its logarithmic
volume derivative $q$ [(e)--(h)], and the Debye temperature
$\thetaD$ [(i)--(l)] as functions of the volume ratio
$\Ve(T)/\Vzero$ along the zero-pressure equilibrium path for diamond
[(a), (e), (i)], MgO [(b), (f), (j)], Si [(c), (g), (k)], and NaCl
[(d), (h), (l)].
Temperature-colored symbols denote values obtained with UMA
(circles), d-UMA (diamonds), UPET(Vinet) (squares), and UPET(AP2)
(pentagons).
The black symbols mark the corresponding values at the static
equilibrium volume $\Vzero$.
The gray curves are guides to the eye.
The horizontal dashed lines in panels (a)--(d) indicate the
representative experimental values shown in
Fig.~\ref{fig:exptgammavsKprime}.
}
\label{fig:gammaqtheta}
\end{figure*}

With the thermal-softening trends in $\Kadiabatic(T)$ and
$\Kisothermal(T)$ reproduced by the present construction with
reasonable, albeit material-, potential-, and EOS-dependent, accuracy,
the underlying $\gamma(V)$ construction can now be assessed through
the behavior of the Gr\"uneisen parameter, its logarithmic volume
derivative, and the Debye temperature along the zero-pressure
equilibrium path, $V=\Ve(T)$, as shown in
Fig.~\ref{fig:gammaqtheta}.
For all four materials, $\thetaD$ decreases monotonically with
increasing volume, consistent with the expected softening of lattice
vibrations upon expansion, while $\gamma$ increases smoothly.
The behavior of $q$, however, depends more strongly on both the
material and the analytical EOS.
For MgO and NaCl, $q$ increases with increasing volume for all four
calculations.
For diamond, UPET(Vinet) gives a decreasing $q$, whereas UPET(AP2)
gives an increasing trend, while for Si the AP2 result is nearly
volume independent and substantially smaller than the Vinet result.
The qualitative behavior of $\gamma(V)$ and $\thetaD(V)$ is therefore
comparatively robust, whereas $q(V)$ is more sensitive to the
analytical EOS.

Relative to the representative experimental estimates indicated by the
horizontal dashed lines in Fig.~\ref{fig:gammaqtheta}(a)--(d), all four
constructions yield larger $\gamma$ values throughout the thermally
sampled zero-pressure paths of the four solids.
The magnitude of this overestimate depends on the material and the
underlying potential: the UPET results lie appreciably closer to the
experimental reference values for diamond and NaCl, whereas d-UMA gives
the largest deviation for Si.
Importantly, the relative ordering of the curves is largely established
by their static values at $V=\Vzero$ and is generally preserved along
the zero-pressure thermal path.
The offsets among the $\gamma$ and $\thetaD$ curves are already apparent
at the static reference state, indicating that they are largely
inherited from the underlying static and elastic description.
In particular, the dispersion correction distinguishing d-UMA from UMA
shifts the elasticity-derived reference values $\thetaDzero$ and
$\gammazero$, together with the equilibrium volume itself.

The comparison between UPET(Vinet) and UPET(AP2) reveals an additional
EOS-dependent effect that cannot be attributed to the underlying static
potential.
The two UPET constructions are anchored to the same static reference
values of $\gammazero$ and $\thetaDzero$, and their $\gamma(V)$ and
$\thetaD(V)$ curves remain comparatively close throughout the thermally
sampled volume range.
Nevertheless, their corresponding $q(V)$ curves may differ
substantially.
For diamond, UPET(AP2) gives slightly larger $q$ values than
UPET(Vinet), whereas for MgO it gives smaller values.
The EOS dependence is more pronounced for Si and NaCl:
UPET(AP2) gives substantially smaller $q$ values than UPET(Vinet) for
Si but substantially larger values for NaCl.
This enhanced sensitivity is expected because
$q=d\ln\gamma/d\ln V$ probes the local volume derivative
of $\gamma$ and can therefore amplify differences between analytical
EOS forms that remain comparatively subtle in $\gamma(V)$ itself.

Accordingly, the present results indicate that both the static-lattice
reference quantities and the volume dependence represented by the
analytical EOS must be described adequately.
Errors in $\thetaDzero$ and $\gammazero$ are propagated from the
underlying static and elastic description, whereas the EOS controls the
subsequent volume dependence and can affect $q$ appreciably even when
the thermal paths of $\gamma$ and $\thetaD$ appear similar.
These discrepancies should not be absorbed into the thermal construction
through fitting to finite-temperature data, since they originate from
limitations of the underlying interatomic potential or the selected
analytical EOS form.

\begin{figure}
  \includegraphics[width=0.47\textwidth]{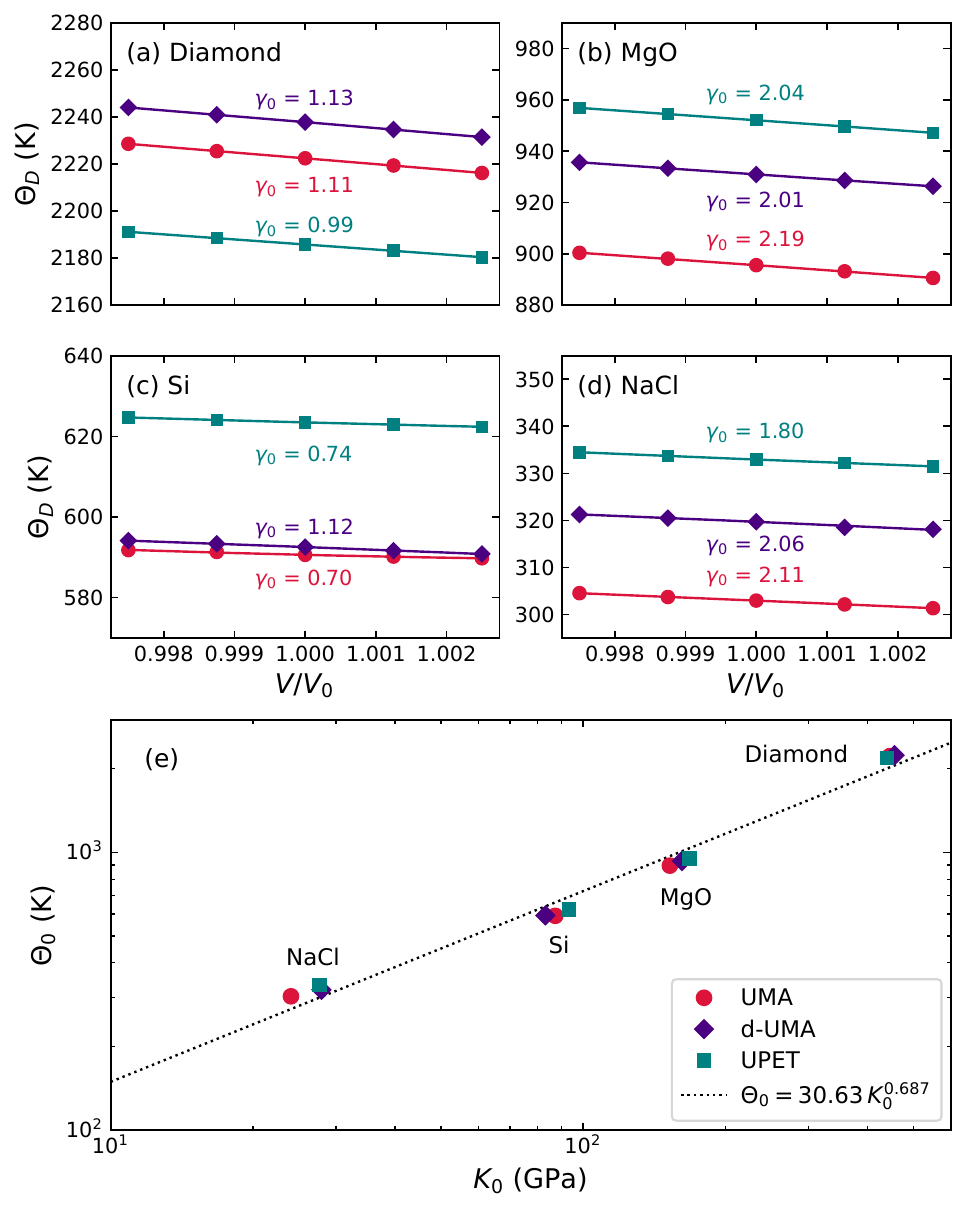}
\caption{
Debye temperatures $\thetaD(V)$ obtained from elastic stiffness tensors
for (a) diamond, (b) MgO, (c) Si, and (d) NaCl, evaluated at five
volumes near $\Vzero$, as predicted by UMA, d-UMA, and UPET.
Symbols denote the calculated values, while solid curves are fits to
the constrained form of Eq.~\eqref{eq:gamma0_fit}, used to extract
$\gammazero$.
(e) Debye temperature at the equilibrium volume, $\thetaDzero$, versus
static bulk modulus $\Kzero$ for the same materials and all three
MLIPs; the dotted line shows the power-law fit to the complete data
set.
}
  \label{fig:thetaDzero}
\end{figure}

Figure~\ref{fig:thetaDzero} illustrates the procedure used to determine
$\gammazero$ from $\thetaD(V)$ by means of the constrained quadratic fit of Eq.~\eqref{eq:gamma0_fit}.
The fit is based on five calculated $\thetaD(V)$ values evaluated at
volumes in a narrow neighborhood of $\Vzero$, over which no mechanical
instability is encountered.
For all materials and MLIPs, the resulting $\thetaD(V)$ curves are
nearly linear over the narrow sampled volume interval.
The constrained quadratic form is nevertheless retained as a local
expansion about $\Vzero$ for the consistent extraction of
$\gammazero$.
The comparison of UMA, d-UMA, and UPET shows that the fitting procedure
remains stable under changes in the underlying MLIP, although the
absolute $\thetaD(V)$ and extracted $\gammazero$ values shift
quantitatively.
The resulting $\gammazero$ values preserve the qualitative separation
between the low-$\gamma$ covalent solids, diamond and Si, and the
higher-$\gamma$ ionic solids, MgO and NaCl.
Figure~\ref{fig:thetaDzero}(e) further shows a power-law correlation
between the Debye temperature at the static equilibrium volume,
$\thetaDzero$, and the static bulk modulus, $\Kzero$, across all four
solids and all three MLIPs.
This trend suggests that inaccuracies in the predicted equilibrium
elastic response are reflected not only in the static bulk modulus but
also in the Debye temperature.

\begin{figure}
  \includegraphics[width=0.47\textwidth]{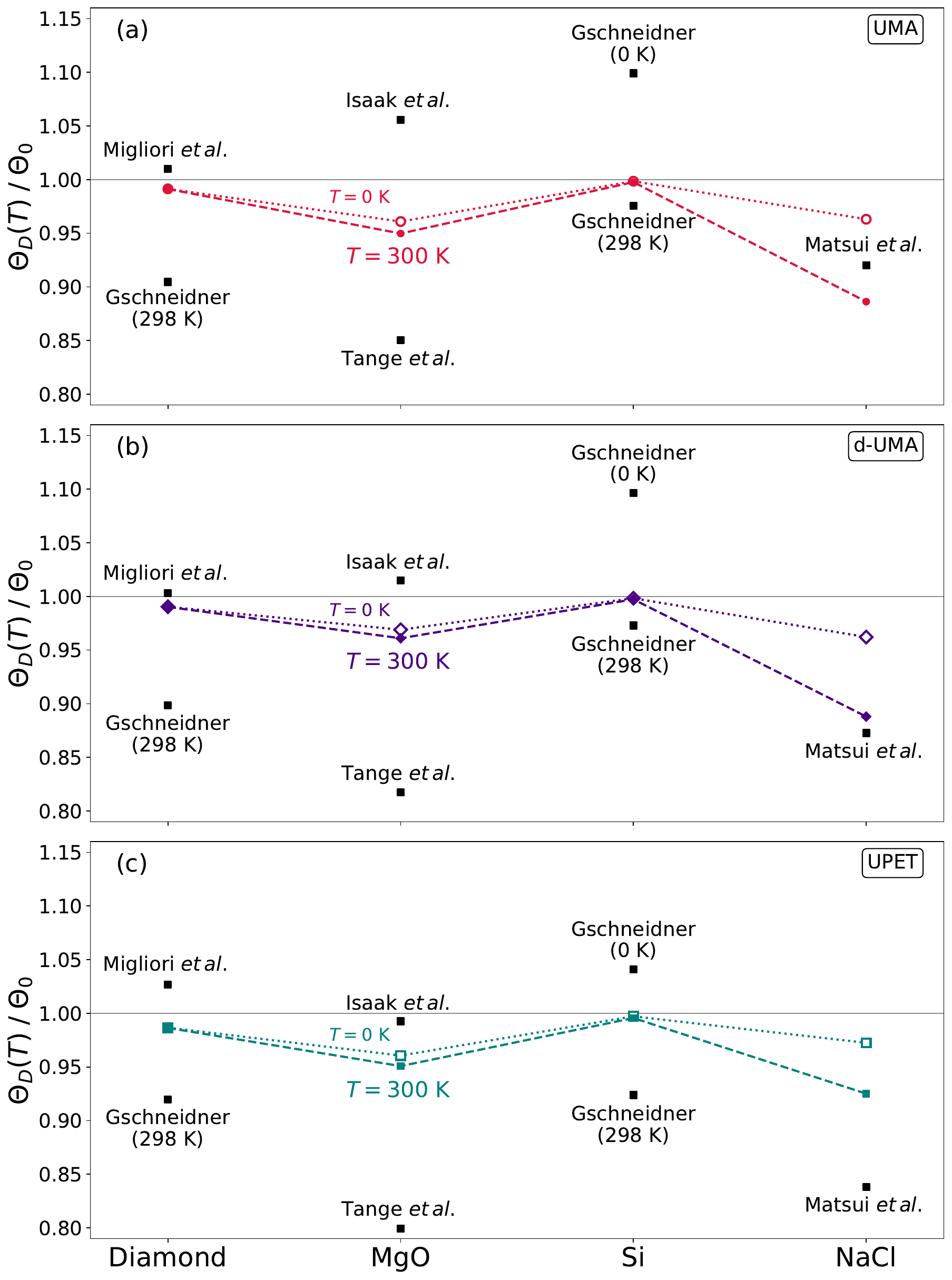}
\caption{
Debye temperatures at $T=0$~K and $300$~K, normalized by the
corresponding static value $\thetaDzero$, from
(a) UMA, (b) d-UMA, and (c) UPET calculations.
Open and filled colored symbols denote the calculated values at
$T=0$~K and $300$~K, respectively.
Solid black squares denote representative experimental $\thetaD(T)$
values
\cite{Migliori2008, gschneidner1964physical, isaak1989measured,
Tange2009, matsui2012simultaneous},
each normalized by the calculated $\thetaDzero$ of the corresponding
MLIP.
The dotted and dashed lines are guides to the eye.
}
  \label{fig:thetaD_scaled}
\end{figure}

Figure~\ref{fig:thetaD_scaled} compares the calculated temperature
dependence of the Debye temperature with representative experimental
values after normalization by the corresponding static value
$\thetaDzero$.
Results are shown separately for UMA, d-UMA, and UPET.
For UPET, the Vinet-based results are shown; the corresponding
AP2-based values are indistinguishable on the scale of the figure.
This normalization reduces the influence of the MLIP-dependent offset
in the absolute Debye temperature and emphasizes its relative thermal
variation.
The calculated values at $T=0$~K need not equal unity because they are
evaluated along the zero-pressure equilibrium path, whereas
$\thetaDzero$ is defined at the static equilibrium volume $\Vzero$.

The calculated Debye temperatures are generally on the same
scale as the representative experimental estimates, although the
experimental values exhibit appreciable scatter.
Where corresponding low- and room-temperature experimental estimates
are available, the calculated reduction in
$\thetaD$ is smaller than that implied by experiment.
The agreement should therefore be interpreted primarily in terms of
the relative material trends rather than as a uniform quantitative
reproduction of the experimental temperature dependence.
For all three MLIPs, diamond and Si remain close to the static reference
value, MgO exhibits an intermediate reduction, and NaCl shows the
largest suppression at $T=300$~K.
This qualitative ordering is common to UMA, d-UMA, and UPET, although
the magnitude of the reduction remains model dependent, most visibly
for NaCl.

\begin{figure}
  \includegraphics[width=0.47\textwidth]{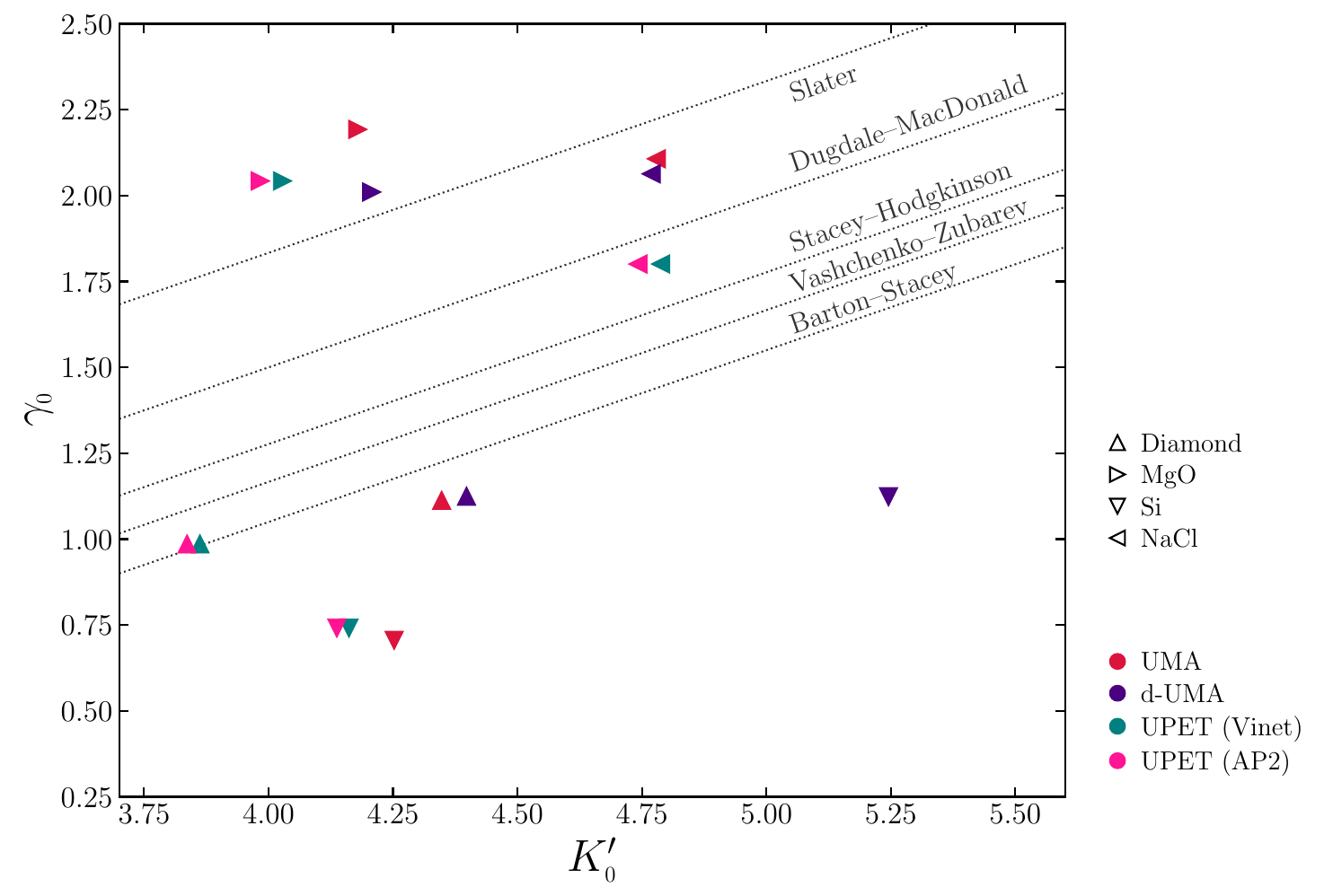}
\caption{
  Elasticity-derived Gr\"uneisen parameter $\gammazero$ versus the
  static-lattice EOS parameter $\Kprimezero$ for the solids considered
  in this work, as obtained from UMA, d-UMA, and UPET.
  For UPET, the $\Kprimezero$ values obtained from the Vinet and AP2
  fits are shown separately.
  The dotted reference lines are drawn using
  $\gammazero = \Kprimezero/2 - 1/6 - \teffzero/3$, as in
  Fig.~\ref{fig:exptgammavsKprime}, with
  $\teffzero=0,1,1.67,2$, and $2.35$ corresponding to the Slater
  \cite{Slater1939}, Dugdale--MacDonald
  \cite{dugdale1953thermal}, Stacey--Hodgkinson
  \cite{StaceyHodgkinson2019}, Vashchenko--Zubarev
  \cite{VashchenkoZubarev1963}, and Barton--Stacey
  \cite{barton1985gruneisen, stacey1995theory} relations,
  respectively.
}
  \label{fig:gamma0_vs_K0p}
\end{figure}

Figure~\ref{fig:gamma0_vs_K0p} revisits the
$\gamma$--$\Kprime$ comparison introduced in
Fig.~\ref{fig:exptgammavsKprime}, but now with $\gammazero$ and
$\Kprimezero$ determined consistently within the present workflow
from elasticity and EOS fitting, respectively.
The same qualitative conclusion emerges: the calculated points do not
collapse onto any single constant-$t$ reference line.
The Si points lie below the conventional constant-$t$ relations,
whereas the diamond points lie near the Barton--Stacey relation, with
the UPET values closest to that line.
MgO lies above the Slater relation, while the NaCl points span the
region between the Slater and Stacey--Hodgkinson relations.
For NaCl, the UMA and d-UMA values lie between the Slater and
Dugdale--MacDonald relations, whereas the UPET values lie between the
Dugdale--MacDonald and Stacey--Hodgkinson relations.
The UPET(Vinet) and UPET(AP2) points lie nearly on top of one another
because they share the same elasticity-derived $\gammazero$, while
their fitted $\Kprimezero$ values differ only slightly.
The addition of AP2 therefore does not materially alter the
$\gammazero$--$\Kprimezero$ comparison itself.
The spread among the different MLIPs reflects variations in both the
elasticity-derived $\gammazero$ and the EOS-derived $\Kprimezero$.

Table~\ref{tab:eos_gruneisen_parameters} complements this comparison
by collecting the parameters of the EOS-based Gr\"uneisen construction
after imposition of the equilibrium anchoring and infinite-compression
constraints.
The coefficient $\tone$ retains an explicit dependence on the
EOS-dependent shift.
Consequently, replacing Vinet by AP2 increases $\tone$ by approximately
one unit for each material, since the fitted values of $\Kprimezero$
change only slightly; see Table~S1 of the Supplemental Material
\cite{supmat}.
The effective equilibrium index $\teffzero$, by contrast, removes this
explicit shift dependence and changes only through the fitted
$\Kprimezero$ when the same elasticity-derived $\gammazero$ is used.
For diamond, the values of $\teffzero$ lie near or slightly above the
Barton--Stacey value of $2.35$, whereas the Si values lie well beyond
the conventional constant-$t$ range.
All MgO values are negative and therefore lie on the opposite side of
the Slater limit from the conventional positive-$t$ relations.
For NaCl, the UMA and d-UMA values lie between the Slater and
Dugdale--MacDonald values, while the UPET values lie between the
Dugdale--MacDonald and Stacey--Hodgkinson values.

Figure~\ref{fig:gamma0_vs_K0p} and
Table~\ref{tab:eos_gruneisen_parameters} therefore establish two
distinct points.
First, the dispersion of the $\gammazero$--$\Kprimezero$ relation
persists when both quantities are obtained consistently from the same
static and elastic input and is thus not merely an artifact of
combining heterogeneous experimental data.
Second, the effective equilibrium index $\teffzero$ is nearly
unchanged between the Vinet and AP2 constructions for the same UPET
input.
By contrast, the sign of $\tone$, which determines the direction of
the volume dependence of the auxiliary function $t(V)$, is not
invariant under the choice of analytical EOS.

\begin{table}
\caption{
Parameters entering the equilibrium anchoring condition
$\gamma(\Vzero)=\gammazero$.
The value of $\gammazero$ is obtained from the fit in
Eq.~\eqref{eq:gamma0_fit}, while $\Kprimezero$ is obtained from the EOS fits in Eqs.~\eqref{eq:Vinet} and \eqref{eq:AP2}.
The parameters $\tone$ and $\teffzero$ are obtained from
Eqs.~\eqref{eq:tone} and \eqref{eq:teffzero}, respectively, using
$\tzero=5/2$ and the corresponding EOS-dependent shift:
$s=1/3$ for Vinet and $s=0$ for AP2.
}
\label{tab:eos_gruneisen_parameters}
\begin{ruledtabular}
\begin{tabular}{@{}lrrlrrr@{}}
Solid & Potential &
\multicolumn{1}{c}{$\gammazero$} &
EOS &
\multicolumn{1}{c}{$\Kprimezero$} &
\multicolumn{1}{c}{$\tone$} &
\multicolumn{1}{c}{$\teffzero$} \\
\hline
Diamond & UMA   & 1.11 & Vinet & 4.35 & -1.18 &  2.68 \\
        & d-UMA & 1.13 & Vinet & 4.40 & -1.22 &  2.72 \\
        & UPET  & 0.99 & Vinet & 3.86 & -0.83 &  2.33 \\
        &       &      & AP2   & 3.84 &  0.21 &  2.29 \\ \ \\
MgO     & UMA   & 2.19 & Vinet & 4.18 &  2.31 & -0.81 \\
        & d-UMA & 2.01 & Vinet & 4.21 &  1.72 & -0.22 \\
        & UPET  & 2.04 & Vinet & 4.03 &  2.09 & -0.59 \\
        &       &      & AP2   & 3.98 &  3.15 & -0.65 \\ \ \\
Si      & UMA   & 0.70 & Vinet & 4.25 & -2.26 &  3.76 \\
        & d-UMA & 1.12 & Vinet & 5.25 & -2.50 &  4.00 \\
        & UPET  & 0.74 & Vinet & 4.16 & -2.02 &  3.52 \\
        &       &      & AP2   & 4.14 & -0.98 &  3.48 \\ \ \\
NaCl    & UMA   & 2.11 & Vinet & 4.78 &  1.15 &  0.35 \\
        & d-UMA & 2.06 & Vinet & 4.77 &  1.04 &  0.46 \\
        & UPET  & 1.80 & Vinet & 4.79 &  0.22 &  1.28 \\
        &       &      & AP2   & 4.74 &  1.29 &  1.21 \\
\end{tabular}
\end{ruledtabular}
\end{table}

\section{\label{s:con}CONCLUSIONS}

An EOS-based construction of the Gr\"uneisen function has been developed and examined through predictions of the finite-temperature isothermal and adiabatic bulk moduli of diamond, MgO, Si, and NaCl.
In this approach, $\gamma(V)$ is constructed from static-lattice EOS information and near-equilibrium elastic data, obtained here from atomistic calculations based on three MLIP-based descriptions, viz., UMA, d-UMA, and UPET, without fitting to experimental thermal data.
The construction combines a volume-dependent auxiliary function $t(V)$, equilibrium anchoring through the elasticity-derived Gr\"uneisen parameter, and enforcement of the infinite-compression limit.
Once an analytic EOS form is adopted, the resulting $\gamma(V)$ is fully determined by the static energy--volume relation and the near-equilibrium elastic input.
The construction has been implemented with both the Vinet and AP2 EOS forms, with the asymptotic shift required for the Vinet construction but not for AP2.

The construction offers a physically coherent route from static atomistic information to the finite-temperature bulk moduli within the Mie--Gr\"uneisen--Debye framework, moving beyond the longstanding constant-$t$ description of $\gamma(V)$ based on a universal $t$ value.
For the four materials considered here, the calculated isothermal and adiabatic bulk moduli exhibit the expected thermal-softening behavior, while the predicted adiabatic bulk moduli reproduce the observed experimental trends and capture the overall scale of $d\Kadiabatic/dT$.
The level of quantitative agreement depends on the material, the underlying interatomic potential, and the selected analytic EOS form.
The comparison among UMA, d-UMA, and UPET shows that discrepancies in the absolute bulk moduli largely reflect the static-lattice baseline supplied by the underlying MLIP.
The comparison between the Vinet and AP2 constructions further shows that the predicted thermal-softening rates exhibit a material-dependent sensitivity to the analytic EOS form, while the leading thermal-softening behavior remains robust.
These results demonstrate that an EOS-based construction of the Gr\"uneisen function can capture the leading thermal-softening behavior of bulk moduli without fitting to thermal data.

The approach developed here requires only static energy--volume data and near-equilibrium elastic information, and can therefore be incorporated directly into workflows based on other MLIPs and empirical potentials that yield sufficiently smooth energy derivatives, as well as into first-principles methods, including DFT.
The observed sensitivity to the underlying static description further suggests that the framework may help identify deficiencies relevant to thermoelastic transferability and thereby inform future training and validation strategies for universal machine-learning interatomic potentials.

The present treatment remains within a quasiharmonic
Mie--Gr\"uneisen--Debye framework and does not include explicit
anharmonic effects beyond this level.
Incorporating explicit anharmonicity will therefore be important in
future extensions.
For applications to lower-symmetry or strongly anisotropic materials,
it will also be necessary to assess the adequacy of reducing the
direction-dependent acoustic response to a single Debye scale and,
where required, to adopt an explicitly anisotropic treatment of the
vibrational spectrum.

\bigskip


\begin{acknowledgments}
Computational workload distribution and execution were managed using PyLauncher \cite{PyLauncher}. 
ChatGPT \cite{ChatGPT} was used as an auxiliary tool for code refactoring, debugging, and optimization of Python routines. 
The author retains sole responsibility for all scientific content, modeling choices, implementation decisions, data validation, interpretation of results, and final verification of the code and reported results.
\end{acknowledgments}

\section*{DATA AVAILABILITY}

The data that support the findings of this article are not publicly available. The data are available from the author upon reasonable request.


\bibliography{refs}

\end{document}